\documentstyle[11pt,newpasp,twoside,epsf]{article}
\newfont{\rmsmall}{cmr10 scaled 900}

\def\edcomment#1{\iffalse\marginpar{\raggedright\sl#1\/}\else\relax\fi}
\reversemarginpar

\begin{document}
\title{The Case for a 30m Diameter Submillimeter Telescope on the Antarctic Plateau}
 \author{Antony A. Stark}
\affil{Smithsonian Astrophysical Observatory, 60 Garden St. MS 12,
Cambridge, MA 02138 USA}

\begin{abstract}
A large single-dish submillimeter-wave telescope equipped with a
focal plane array containing $\sim10^4$ bolometers 
and costing about $\$120{\mathrm{M}}$ could locate
most protogalaxies in the southern sky within a year of
operation. 
\end{abstract}

Many of the telescopes planned for the next few decades are
designed to observe high-redshift galaxies in the process of
formation (NRC 2001).
These instruments, such as the James Webb Space Telescope (JWST), 
the OverWhelmingly Large Telescope (OWL), and the 
Atacama Large Millimeter Array (ALMA), will
have sufficient sensitivity and resolution to observe 
detailed structure within protogalaxies; they will not,
however, have sufficient field of view to survey large
areas of sky and discover objects to study.
Consider, for example, the ALMA.  
As seen in Figure 1,
protogalaxies typically have a flux density 
at $\lambda450\micron$ which is
$\la 10$ mJy.
The ALMA can detect such a source in 3 minutes of observing time. 
That's really fast.
The size of an ALMA map, however, is $\sim 2 \times 10^{-5}$ square degree,
so to survey a square degree at this sensitivity would require
$\sim 100$ days.   If the ALMA were dedicated to a sky survey 
for ten years, it would be able to cover about $10^{-3}$ of the
entire sky.

A 30-m submillimeter-wave telescope operating on the Antarctic Plateau
with a focal-plane array of bolometers would be able
to survey the southernmost \onethird \, of the sky in a year.
Observatory sites on the Antarctic Plateau have exceptional
submillimeter-wave sky transparency and stability (Chamberlin 2001, 
Peterson et al. 2003).  Technological
progress in submillimeter-wave detectors will make possible
focal-plane arrays containing many thousands of bolometers.
A 10-meter class single-dish telescope designed for
such arrays and located at the South Pole has been approved, and
construction is expected to begin this year (Stark 2003).  Looking ahead, a
30-meter class telescope could be made sufficiently
accurate for submillimeter and far-infrared work through a modest
application of active surface techniques.  A basic design similar to the
IRAM 30-m could be combined with crude active control of the primary
mirror panel
alignment for a total cost of 
$\sim \$120$M.  The 2\arcsec \, to 5\arcsec \,
beamsize of such a telescope would be well-coupled to protogalaxies.
With a field of view $\sim 10^{-1}$ square degree in size,
this instrument could survey the entire sky south of $\delta \approx -25\deg$ with
1~mJy sensitivity in a year.   Almost all
protogalaxies and protostellar cores would be found, and could
be distinguished on the basis of $\lambda350\micron$ to $\lambda450\micron$ 
color.  The resulting catalog
would be a treasure trove of objects for high-resolution study with the giant
telescopes to come.

\begin{figure}[t!]
\plotfiddle{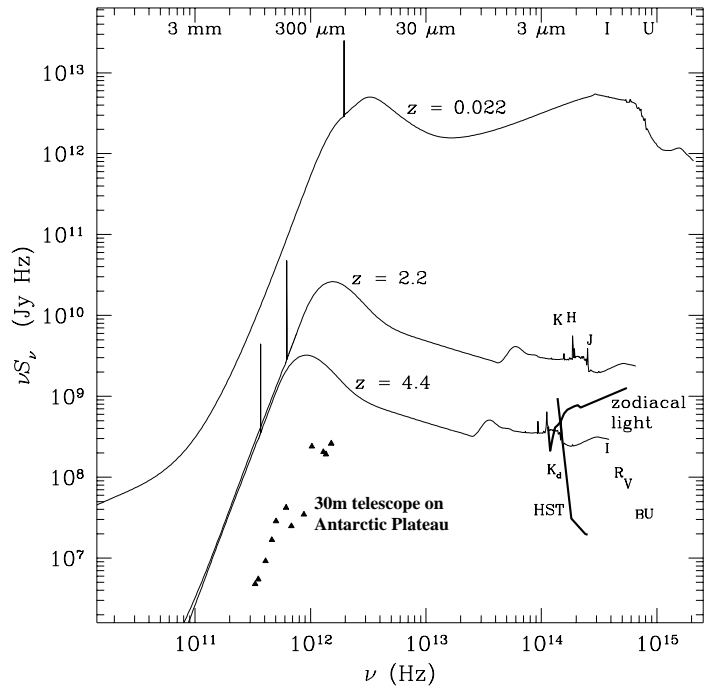}{3.15in}{0}{86}{86}{-155}{-407}
\caption{
Normal galaxies at low and high redshift.  The broad-band galaxy spectrum
labelled $z=0.022$ has the luminosity and spectrum of M99, a normal $L^*$
spiral.  The spectra labeled $z=2.2$ and $z=4.4$ are models of the 
initial starburst
in such a galaxy at two possible eras, evolved in a
standard CDM model (Katz 1992) with $h_{75} = 1$ and $\Omega=1$.
The 158\micron \, $\rm C^+$ line is shown to scale; other lines are suppressed.
The points KHJIRVBU are 1\% of the sky brightness in a square arcsecond at
Mauna Kea (CFHT Observer's Handbook).  The point $\rm K_d$ is 1\% of the
sky brightness in a square arcsecond at the South Pole at $\lambda 2.3 \mu \rm m
$.
The curve labelled HST is the limiting sensitivity of the NICMOS on the
Hubble Space Telescope.
The triangles show the continuum sensitivity in one hour of a 30
meter Antarctic telescope in the submillimeter-wave atmospheric windows
(Stark 1997).
}
\end{figure}

\acknowledgments
Support was provided by NSF grant OPP-0130612.

\end{document}